\documentclass[11pt,a4paper]{article} 
\usepackage{jheppub}

\usepackage{latexsym}
\usepackage{amssymb}

\numberwithin{equation}{section}

\newcommand{\be}{\begin{equation}}
\newcommand{\ee}{\end{equation}}
\newcommand{\bea}{\begin{eqnarray}}
\newcommand{\eea}{\end{eqnarray}}
\newcommand{\bean}{\begin{eqnarray*}}
\newcommand{\eean}{\end{eqnarray*}}
\newcommand{\nn}{\nonumber}
\newcommand{\bra}{\langle}
\newcommand{\ket}{\rangle}
\newcommand{\eps}{\epsilon}
\newcommand{\tr}{\mbox{tr}\,}
\newcommand{\Tr}{\mbox{Tr}\,}

\newcommand{\vecx}{{\mathbf x}}

\newcommand{\rmR}{{\rm R}}
\newcommand{\rmI}{{\rm I}}

\newcommand{\half}{\frac{1}{2}}

\newcommand{\C}{\mathbb{C}}
\newcommand{\R}{\mathbb{R}}

\newcommand{\cH}{{\cal H}}
\newcommand{\cM}{{\cal M}}
\newcommand{\cO}{{\cal O}}

\title{Stability of complex Langevin dynamics in effective models}

\author[a]{Gert Aarts,}
\author[b]{Frank A.\ James,}
\author[c]{Jan M.\ Pawlowski,}  
\author[d]{Erhard Seiler,}
\author[c]{D\'enes Sexty}
\author[c,e]{and Ion-Olimpiu Stamatescu}
 
\affiliation[a]{Department of Physics, College of Science, Swansea University, Swansea, United Kingdom}
\affiliation[b]{Institute for Theoretical Physics, Universit\"at Regensburg, Regensburg, Germany} 
\affiliation[c]{Institut f\"ur Theoretische Physik, Universit\"at Heidelberg, Heidelberg, Germany}
\affiliation[d]{Max-Planck-Institut f\"ur Physik (Werner-Heisenberg-Institut), M\"unchen, Germany}
\affiliation[e]{FEST, Heidelberg, Germany}

\emailAdd{g.aarts@swan.ac.uk}
\emailAdd{Frank.James@physik.uni-regensburg.de} 
\emailAdd{J.Pawlowski@thphys.uni-heidelberg.de}
\emailAdd{ehs@mppmu.mpg.de}
\emailAdd{d.sexty@thphys.uni-heidelberg.de}
\emailAdd{i.o.stamatescu@thphys.uni-heidelberg.de}

\abstract{
 The sign problem at nonzero chemical potential prohibits the use of 
importance sampling in lattice simulations. Since complex Langevin 
dynamics does not rely on importance sampling, it provides a potential 
solution.  Recently it was shown that complex Langevin dynamics fails in 
the disordered phase in the case of the three-dimensional XY model, while 
it appears to work in the entire phase diagram in the case of the 
three-dimensional SU(3) spin model.  Here we analyse this difference  
and argue that it is due to the presence of the nontrivial Haar measure in the 
SU(3) case, which has a stabilizing effect on the complexified dynamics. 
The freedom to modify and stabilize the complex Langevin process is discussed in some detail. 
 }
 

\arxivnumber{1212.5231}

\begin{document}
\maketitle


\section{Introduction}
\label{sec:intro}

Because the fermion determinant in QCD at nonzero baryon chemical potential is complex, standard lattice QCD algorithms based on importance sampling cannot be used. As a result, a nonperturbative study of the QCD phase structure in the temperature -- chemical potential plane is still missing \cite{pdf}. 
Complex Langevin dynamics \cite{Parisi:1984cs,Klauder:1983} may provide a solution, since it is not based on importance sampling, but instead on a stochastic exploration of an enlarged (complexified) field space. However, the method is not guaranteed to work:
the numerical solution of a complex Langevin process may converge to a wrong answer. 
This problem was observed immediately \cite{Ambjorn:1985iw,Ambjorn:1986fz}  after complex Langevin dynamics was proposed in the early 1980s and is still present in current complex Langevin simulations \cite{Berges:2006xc,Berges:2007nr,arXiv:1005.3468}. However, recently it has also been shown convincingly that in some cases complex Langevin simulations converge to the seemingly correct answer, even when the sign problem is severe, i.e.\ in the thermodynamic limit of one-, three- and four-dimensional field theories at nonzero chemical potential \cite{Aarts:2008wh,Aarts:2009hn,arXiv:1006.0332,Aarts:2011zn}. Given the importance of understanding strongly-coupled theories such as QCD at nonzero chemical potential numerically, this issue clearly needs to be addressed.

The outstanding questions are therefore  (1) to quantify whether the results from complex Langevin simulations have converged correctly; (2) to understand why the (in)correct convergence occurs; (3) to find a cure in the case of incorrect convergence.
Recently we have clarified question (1) in some detail by putting the mathematical foundation of complex Langevin dynamics on firmer footing and deriving a set of criteria for correctness that need to be satisfied \cite{arXiv:0912.3360,arXiv:1101.3270}. These consistency conditions can be calculated during the complex Langevin simulation.

 In this paper we focus on question (2), yielding insight that can be used to address the third question.
This is done in the context of two three-dimensional spin models recently studied:  the abelian XY model and the nonabelian SU(3) spin model, both at nonzero chemical potential. The XY model was studied in ref.~\cite{arXiv:1005.3468}. Here it was shown that correct convergence is obtained deep in the ordered phase, but incorrect convergence was found in the disordered phase and the transition region. This conclusion was based on a study of the expected analyticity of observables around $\mu^2=0$, of properties of distributions in the complexified field space, and from a comparison with an alternative world-line formulation, which is sign-problem free \cite{arXiv:1001.3648}.
 The SU(3) spin model was studied in ref.\ \cite{Aarts:2011zn}, extending the classic papers \cite{KW,BGS}. Using similar criteria as in the XY model (analyticity and Taylor series expansion around $\mu^2=0$, localized distributions in the complexified field space) as well as a test of the criteria for correctness mentioned above, we concluded that complex Langevin dynamics yields the correct results in the entire phase diagram. This conclusion was subsequently supported by a study using a dual formulation, which is again sign-problem free \cite{arXiv:1104.2503,Mercado:2012ue}.

In the present paper we address this observed difference in performance of the complex Langevin method between the XY and the SU(3) spin model. Surprisingly, we find that the nonabelian nature of the SU(3) spin model is crucial. In particular we demonstrate that in the disordered phase of the XY model the real manifold is unstable against complex fluctuations while in the SU(3) spin model it is stable. Stability of the real manifold is important for understanding the expected analyticity of observables around $\mu^2=0$.

The paper is organized as follows. In sec.~\ref{sec:3d} we remind the reader of the SU(3) and XY models and reduce them to effective one-link models with complex couplings. In sec.~\ref{sec:3} the effective one-link models are discussed in detail  and the difference between the abelian and nonabelian models is stressed. In sec.~\ref{sec:trafo} we use the insight to illustrate how coordinate transformations may  stabilize the dynamics. Conclusions and an outlook are given in ref.~\ref{sec:conclusion}. Appendix \ref{sec:app} contains a detailed discussion of possible modifications of the complex Langevin approach.

\section{SU(3) and U(1)  spin models}
\label{sec:3d}
  
\subsection{SU(3) spin model}
\label{sec:su3}

We consider the three-dimensional SU(3) spin model at nonzero chemical 
potential, with the action
\be
 S = S_B+S_F,
 \ee
 where
 \begin{align}
 S_B  & = -\beta\sum_{x}\sum_{\nu=1}^3 \left(  \Tr U_x\Tr 
 U_{x+\hat\nu}^\dagger + \Tr U_x^\dagger\Tr U_{x+\hat\nu}\right),
 \\
 \label{eq:sf}
S_F & =  -h \sum_x\left( e^\mu\Tr U_x + e^{-\mu}\Tr U_x^\dagger\right).
\end{align}
The matrices $U_x$ are elements of SU(3). The action is complex and satisfies the usual symmetry, $S^*(\mu) = S(-\mu^*)$. The `fermion' contribution follows from the full QCD determinant by preserving only propagation in the temporal direction at leading order in the hopping expansion (i.e.\ heavy quarks).
The model is part of a whole family of effective Polyakov loop models with heavy quarks \cite{Fromm:2011qi,Fromm:2012eb}.
 A detailed complex Langevin study  can be found in ref.\ \cite{Aarts:2011zn}. 

We want to construct a one-link model that captures the essential dynamics.
We therefore focus on a site $x$ and consider the interaction of $U_x$ with its nearest neighbours. Treating all six neighbours equally and denoting their contribution as $u\in\mathbb C$, we write the effective one-link model as\footnote{As always, $U^{-1}$ is written instead of $U^\dagger$ to allow for the correct extension to SL(3,$\mathbb C$).} 
\be
S_U = - \beta_1 \Tr U - \beta_2\Tr U^{-1},
\ee
where the effect of the neighbours is captured,  to a certain extent,  by a
simplified parametrization 
\be
 \beta_1 = 6\beta u +h e^\mu, 
 \quad\quad\quad
  \beta_2 = 6\beta u^* +h e^{-\mu}, 
\ee
with $\beta^*_1(\mu) = \beta_2(-\mu^*)$.

\begin{figure}
  \centerline{
    \includegraphics[width=0.8\textwidth]{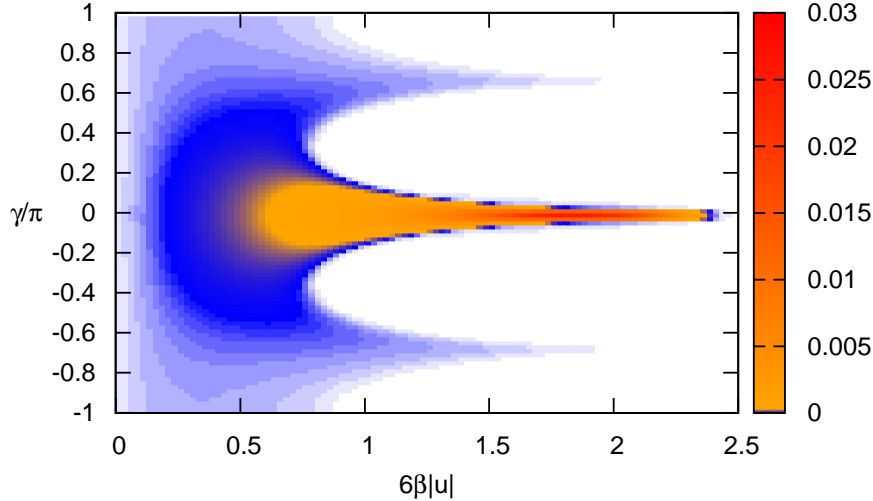}
  }
 \caption{
 Correlation between the phase $\gamma$ and the modulus $6\beta|u|$, where $u =  |u|e^{i\gamma}
 =\Tr U$ in the three-dimensional SU(3) spin model, for 15 combinations of
$\beta=0.125, 0.13, 0.135$, and $\mu=0.5, 1,2,3,4$, at $h=0.02$ on a $12^3$ lattice.
 }
\label{fig:hist}
 \end{figure}

Typical values of $\beta_{1,2}$ are determined by $\beta, h$ and $\mu$, and by the contribution from the nearest neighbours, represented by $u$, in the original three-dimensional theory. 
We first note that the critical $\beta$ value in the SU(3) spin model at vanishing $\mu$ is $\beta_c\sim 0.133$ \cite{Aarts:2011zn,Mercado:2012ue}.\footnote{Its relation with the four-dimensional coupling can be ultimately be understood from a combined strong-coupling/hopping parameter  expansion \cite{Fromm:2011qi}.}  We therefore consider $\beta$ values around this value, as in refs.\ \cite{Aarts:2011zn,Mercado:2012ue}.
Writing $u=|u| e^{i\gamma}$, the correlation between its phase and amplitude can be determined using simulations in the full, three-dimensional theory. The result 
is shown in Fig.\ \ref{fig:hist}, from a study on a $12^3$ lattice for 15 combinations of $\beta$ and $\mu$ at fixed $h=0.02$. Note that during a Langevin simulation, $u$ takes values in the familiar triangular shape with corners at 
$3e^{2\pi i q/3}$ ($q=0,1,2$). This explains the three spikes at $\gamma/\pi = 0, \pm 2/3$. At larger $\beta$ and/or $\mu$, $u$ lies predominantly in the trivial direction ($q=0$) and $\gamma$ is closer to zero. 
Note also that at nonzero $\mu$, the dynamics takes place slightly outside SU(3) \cite{Aarts:2011zn} and $|u|$ is not strictly bounded by~3. Typical values of $h$ and $\mu$ are determined by the relation with four-dimensional theory \cite{Aarts:2008rr}: for Wilson quarks, $h=(2\kappa)^{N_\tau}$ with $\kappa\sim 0.12$ and $N_\tau=4, 6, 8$, etc. Relevant values of $h$ are therefore small.
Upon identification with the four-dimensional theory, $\mu$ as used here 
in fact  corresponds to $\mu/T$ (with $1/T=a_\tau N_\tau$).

We finally express the complex couplings as
\be
 \beta_1 = \beta_{\rm eff} e^{i\gamma} +h e^\mu, 
 \quad\quad\quad
  \beta_2 =  \beta_{\rm eff} e^{-i\gamma} +h e^{-\mu}, 
\ee
where $\beta_{\rm eff} = 6\beta|u| \lesssim 2.5$.

\subsection{XY model}
\label{sec:xy}

The action for the three-dimensional XY model at nonzero chemical potential reads
\bea
S &=&  -\frac{\beta}{2}\sum_x\sum_{\nu=1}^3 \left( e^{\mu\delta_{\nu,3}} U_x 
U_{x+\hat\nu}^\dagger + e^{-\mu\delta_{\nu,3}}U_x^\dagger U_{x+\hat\nu} \right) 
\nn \\
&=&  -\beta\sum_x\sum_\nu\cos\left(\phi_x-\phi_{x+\hat\nu}
-i\mu\delta_{\nu,3}\right),
\eea
where $U_x = e^{i\phi_x}$ are in this case U(1) phase variables. A complex Langevin study of this model can be found in ref.\ \cite{arXiv:1005.3468}.

We first note that deep in the ordered phase, where $\phi_x-\phi_{x+\hat\nu}\ll 1$, 
the action reduces to 
\be
S \sim \half\sum_{x,\nu} \beta_{\nu}\left( \phi_x-\phi_{x+\hat\nu}\right)^2,
\ee
with $\beta_{1,2}=\beta$, $\beta_3=\beta\cosh\mu$.
Here we dropped an overall constant and employed the periodicity of the lattice. Since
this action is real, the sign problem is absent in this limit. Indeed, in the corresponding part of the phase diagram (large $\beta$ and/or $\mu$), no problems were encountered in ref.\ \cite{arXiv:1005.3468}.

As in the SU(3) case, we consider an effective one-link model of the form
\be
S_U = -\beta_1 U - \beta_2 U^{-1}, \quad\quad\quad U=e^{i\phi},
\ee
with complex couplings $\beta_{1,2}$. If we follow the approach from the SU(3) model and replace the six nearest neighbours by a common contribution $u\in\mathbb C$,  we find that
\be
\beta_1  = \beta u^*\left(2+\cosh \mu\right), 
\quad\quad\quad
\beta_2 = \beta u    \left(2+\cosh \mu\right),
\ee
so that $\beta_1^*=\beta_2$ and the action $S_U$ is in fact real. Hence we are forced to represent the nearest neighbours by independent complex phases and we consider the general effective U(1) one-link model,
\be
S_U = -\beta_1 U - \beta_2 U^{-1} = -\beta'_1\cos \phi - \beta'_2\sin\phi,
\ee
where we take $\beta_{1,2}, \beta'_{1,2}\in \mathbb C$ without further restrictions.

\section{Complex Langevin for effective models}
\label{sec:3}

\subsection{U(1) one-link model}

We consider complex Langevin dynamics for the one-link models constructed above. In the U(1) case, the partition function reads
\be
\label{eq:u1}
Z_{{\rm U}(1)} =  \int_{-\pi}^\pi d\phi\, e^{-S_U(\phi)},
\ee
and the corresponding complex Langevin equation is
\be
\dot\phi  = K[\phi(t)] +\eta(t), 
\quad\quad\quad K(\phi) = -\frac{\partial S_U}{\partial\phi},
\ee
with standard relations for the noise, $\bra\eta(t)\ket=0$,  $\bra\eta(t)\eta(t')\ket=2\delta(t-t')$. Since the action and the drift term $K$ are complex, $\phi$ will take values in the complex plane and $U=e^{i\phi}$ is no longer a phase variable.

\subsection{SU($N$) one-link models}
\label{sec:sun}

In the case of the SU(3) [or more generally an SU($N$)] spin model, we can express all dynamics in terms of the eigenvalues of $U$, subject to the constraints, i.e.,
\be
\label{eq:diag}
U  = \mbox{diag}\left(e^{i\phi_1}, e^{i\phi_2}, \ldots, e^{i\phi_N} \right),  \quad\quad\quad 
\phi_1+\phi_2+\ldots+\phi_N = 0.
\ee
The partition function then includes integrating over the reduced Haar measure, which represents explicitly the integration over the full group manifold,
\be
H(\{\phi_a\})  = \prod_{a<b} \sin^2\left(\frac{\phi_a-\phi_b}{2}\right),
\ee
($a,b=1,\ldots,N$)  and reads
\bea
Z_{{\rm SU}(N)} &=&  \int_{-\pi}^\pi d\phi_1\ldots d\phi_N\, \delta\left(\phi_1+\phi_2+\ldots+\phi_N\right) H(\{\phi_a\}) e^{-S_U(\{\phi_a\})}
\nn \\
&=&  \int_{-\pi}^\pi d\phi_1\ldots d\phi_N\, \delta\left(\phi_1+\phi_2+\ldots+\phi_N\right) e^{-S_{\rm eff}(\{\phi_a\})},
\eea
with 
\be
 S_{\rm eff} = S_U+S_H = S_U - \ln H.
 \ee
 The non-abelian Haar measure, which is explicitly seen here as the reduced Haar measure, is the crucial difference between  SU($N$) and  U(1) spin models. 
This is especially clear in the case of SU(2) with a simple action
\be
S_U = -\frac{\beta}{2}\Tr U = -\beta\cos\phi, \quad\quad\quad \beta\in\mathbb C.
\ee
As in the U(1) case, the partition function involves a single integral over $\phi\equiv\phi_1=-\phi_2$ only and reads
\be
Z_{{\rm SU}(2)} = \int_{-\pi}^\pi d\phi\, \sin^2(\phi) e^{-S_U(\phi)}.
\ee
The presence of the reduced Haar measure is the important difference with Eq.~(\ref{eq:u1}).

To write down the Langevin equations in the SU($N$) case, we may follow several routes. Firstly, we may  eliminate $\phi_N$ using the constraint in Eq.\ (\ref{eq:diag}), and write Langevin equations for the remaining $N-1$ degrees of freedom, 
\be
\dot\phi_a  = K_a[\phi(t)] +\eta_a(t), 
\quad\quad\quad K_a(\phi) = -\frac{\partial S_{\rm eff}}{\partial\phi_a}.
\ee
Each update requires $N-1$ stochastic kicks. This is the approach followed in refs.\ \cite{KW,Aarts:2011zn}.

Secondly, the constraint can also be implemented by introducing new variables $z_a$ 
($a=1, \ldots, N$), according to
\be
\label{eqz}
\phi_a = z_a - \frac{1}{N}\left(z_1+z_2+\ldots +z_N\right), 
\ee
such that the constraint $\sum_a\phi_a=0$ is automatically satisfied.
This is a singular transformation; the validity of the procedure 
therefore requires some justification, which is given in Appendix 
\ref{singlin}. It should be noted that the stochastic process, even when 
restricted to the first $N-1$ variables $\phi_a$, is different from the 
one above.
We then write the following dynamics for $z_a$,
\be
 \dot z_a  = K_a[z(t)] +\eta_a,   \quad\quad\quad K_a(z) = -\frac{\partial S_{\rm eff}}{\partial z_a},
\ee
where the $\phi_a$ are always considered as functions of the $z_a$, according to Eq.~(\ref{eqz}). 
 In this formulation $N$ stochastic variables are used. However, it is easy to see that 
the real part of the centre-of-mass coordinate is freely diffusing (the imaginary part is not evolving at all), since
\be
\sum_a K_a(z) = 0 \quad\quad \Rightarrow \quad\quad \sum_a   \dot z_a = \sum_a \eta_a.
\ee 
Since the real parts are angular degrees of freedom, this does not cause 
harm in terms of runaways.

Finally, we may also write the dynamics directly for the SU($N$) matrix $U$ and not in terms of the eigenvalues. 
After discretizing the Langevin time with stepsize $\eps$, this takes the
form
\be
\label{eq:313}
U(t+\eps) = \exp\left[ i\lambda_a\left(\eps K_a+\sqrt\eps\eta_a\right)\right] U(t),
\ee
where $\lambda_a$ (here $a=1,\ldots, N^2-1$) are the traceless, hermitian 
Gell-Mann matrices, normalized as $\Tr\lambda_a\lambda_b=2\delta_{ab}$
and the drift terms are
\be
K_a = -D_aS_U = i\beta_1\Tr\lambda_aU - i \beta_2\Tr\lambda_aU^{-1}.
\ee 
Note that the complex process runs in the SL($N,\mathbb{C}$) extension of SU($N$)
and thus, e.g.\ in Eq.\ (\ref{eq:313}),  $U \in$  SL($N,\mathbb{C}$) and $K_a \in$ sl($N, \mathbb{C}$).
In this formulation there are $N^2-1$ stochastic variables, instead of $N-1$. In ref.\ \cite{Berges:2007nr} it was shown how the additional noise variables generate the correct drift for the remaining degrees of freedom when $U$ is diagonalized after each update. This formulation was considered  in refs.\ \cite{BGS,Berges:2006xc,Berges:2007nr,Aarts:2008rr}.

\subsection{Classical flow}

The role of the reduced Haar measure can be seen by studying the classical flow diagrams. 
Since the classical flow for the SU(3) spin model with angles $\phi_{1,2}$ takes place in four real 
dimensions, it is hard to give a proper graphical representation. Instead we 
present as an illustration the classical flow for the SU(2) one-link model, in which the reduced Haar measure plays a similar stabilizing role. 
  
Explicitly, we compare the U(1) and SU(2) case, with
\be
\label{eq:usu}
Z_{{\rm U}(1)} = \int_{-\pi}^\pi dx\, e^{\beta\cos x}, 
\quad\quad\quad
Z_{{\rm SU}(2)} = \int_{-\pi}^\pi dx\, \sin^2 x\,e^{\beta\cos x},
\ee
with complex $\beta=\beta_\rmR+i\beta_\rmI$. The real and imaginary parts of the drift $K$, after complexification $x\to x+iy$, are
\bea
K_\rmR 
 &=& -\beta_\rmR\sin x\cosh y+\beta_\rmI\cos x\sinh y - \frac{2d\sin(2x)}{\cos(2x)-\cosh(2y)}, 
\\
K_\rmI 
 &=& -\beta_\rmR\cos x\sinh y+\beta_\rmI\sin x\cosh y + \frac{2d\sinh(2y)}{\cos(2x)-\cosh(2y)}.
\eea
In the U(1) case, the contribution from the Haar measure is absent and $d=0$; in SU(2), $d=1$.

\begin{figure}
\centerline{
  \includegraphics[width= 0.45\textwidth]{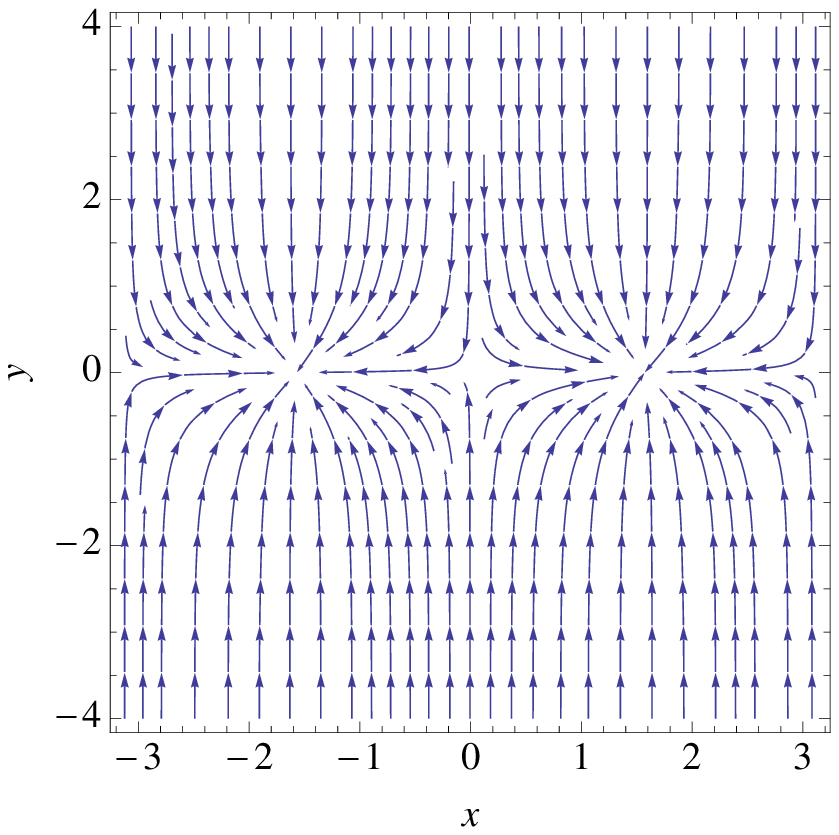}
  \includegraphics[width=0.45\textwidth]{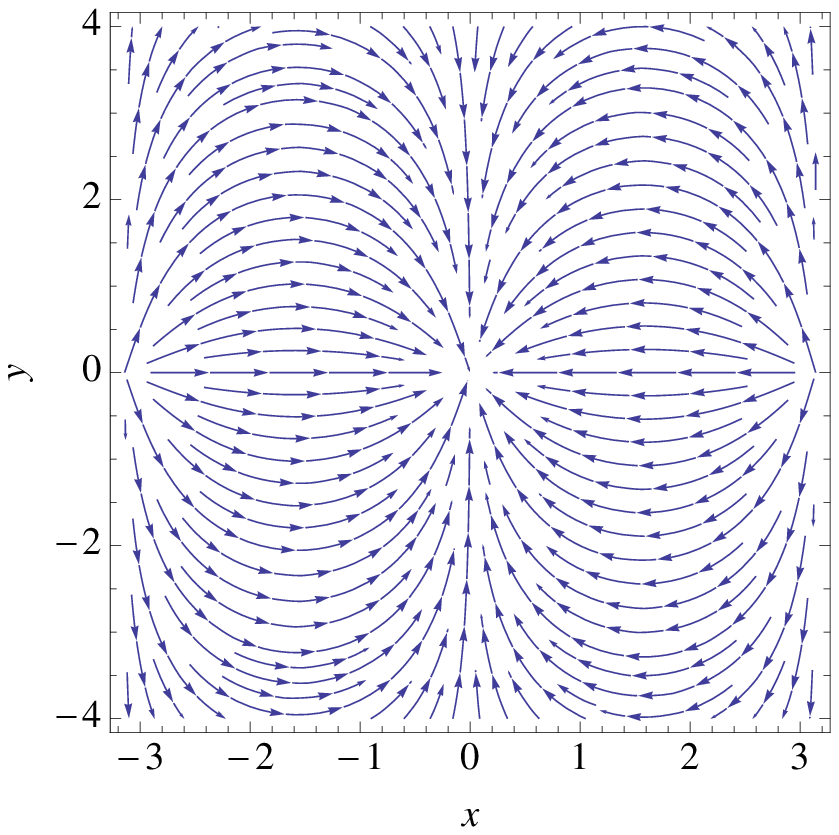} }
  \caption{Classical flow diagrams: drift from the Haar 
   measure only, in SU(2) (left);  drift from the action only, in U(1) and SU(2), with $\beta=1$ (right).
   }
\label{fig:class1}
\end{figure}

 Let us first consider the contribution from the Haar measure.  It is easy to see that when $|y|$ is large,
 \be
  K_\rmR \to 0, \quad\quad\quad K_\rmI = -2\mbox{sgn}(y)
  \quad\quad\quad\quad \mbox{(Haar measure only)},
\ee 
 i.e.\ the contribution is purely restoring. 
  We also note that at the origin and at ($\pm\pi,0$) the drift force from the Haar measure becomes singular; this 
invalidates potentially the formal argument for correctness of the complex Langevin approach 
method, because it might produce boundary terms at the singularity. 
Practically, however, this does not seem to cause problems, as the 
equilibrium measure appears to vanish sufficiently strongly at the origin (although it does require the use of an adaptive stepsize algorithm in simulations \cite{arXiv:0912.0617}). 
Finally, there are stable fixed points at $(x,y)=(\pm\pi/2,0)$. 
  These findings are illustrated in Fig.~\ref{fig:class1} (left), where the drift from the Haar measure only is shown.

Let us now consider the contribution from the action by itself, i.e.\ the drift in the U(1) model ($d=0$), see 
Fig.\ \ref{fig:class1} (right). This drift has an attractive fixed point at the origin and repulsive fixed points at $(\pm\pi,0)$. Due to these repulsive fixed points, the real manifold ($y=0$) is unstable against small fluctuations in the $y$ direction. 
Therefore even with stochastic kicks in the $x$ direction only (real noise), fluctuations in the imaginary direction, e.g.\ due to a small but nonzero $\beta_\rmI$,  will immediately lead to an exploration of the complexified space, resulting in trajectories that may take large excursions in the $y$ direction, 
due to the character of the repulsive flow around $x=\pm\pi$.

\begin{figure}
 \centerline{
 \includegraphics[width= 0.45\textwidth]{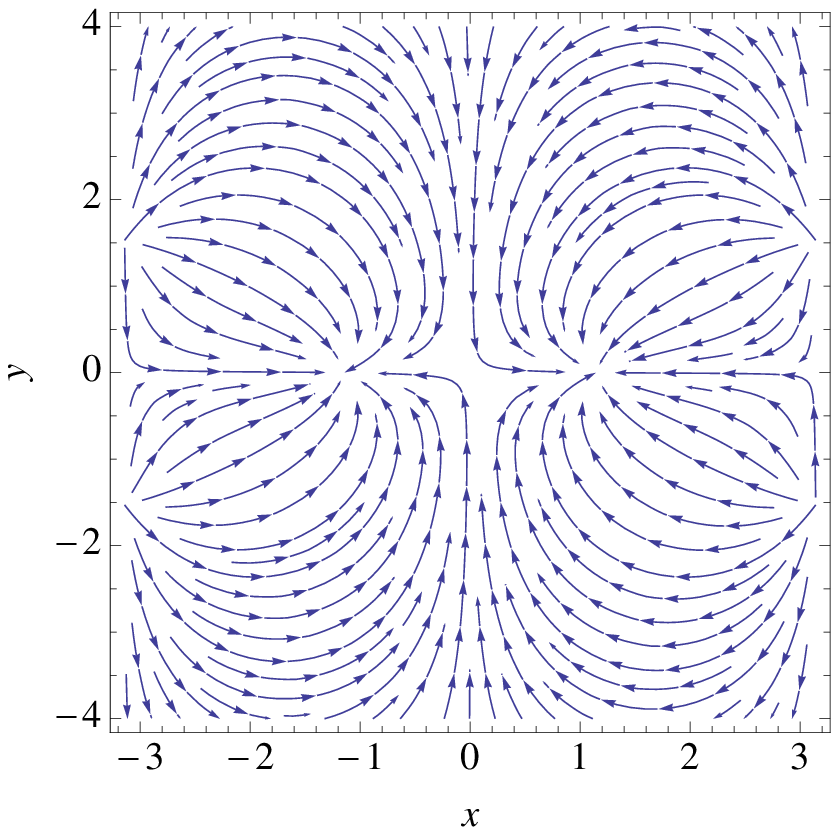}
 \includegraphics[width=0.45\textwidth]{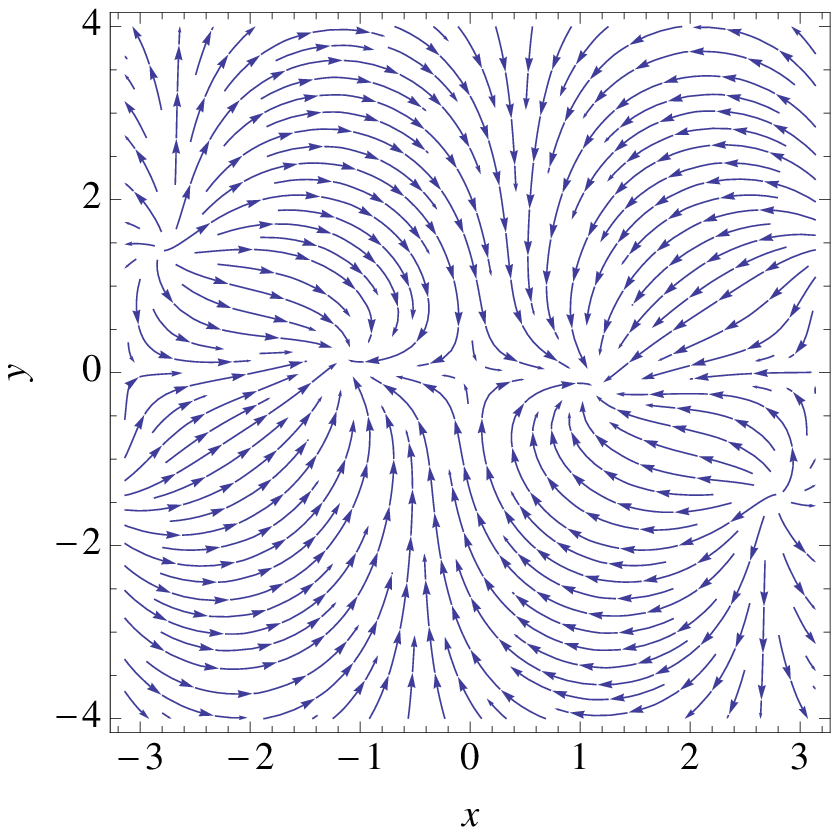}
  }
 \caption{Classical flow diagrams for SU(2) for $\beta=1$ (left) and $\beta=1+0.5i$ (right).
 }
 \label{fig:class2}
\end{figure}

In the nonabelian case, one has to add the contribution of the Haar measure to that of the action.
The resulting flow is shown in Fig.\ \ref{fig:class2} (left) for real $\beta=1$. The singular flow at the origin and at  ($\pm\pi,0$) remains.
We observe that the repulsive fixed points have moved away from the real manifold to $(\pm\pi, \pm y_*)$, while the attractive fixed points remain on the real manifold at ($x_*,0$), where
\be
\cos x_* = -\frac{d}{\beta}+ \sqrt{1+\frac{d^2}{\beta^2}},
\quad\quad\quad
\cosh y_* = \frac{d}{\beta}+\sqrt{1+\frac{d^2}{\beta^2}}, 
\ee
(for $\beta=1$ this yields $x_*=\pm 1.14$, $y_*=\pm 1.53$). Note that the presence of the reduced Haar measure ($d=1$) is essential. Since the repulsive fixed points have moved away, the real axis is stable in SU(2) for $\beta$ values that are not too large. This can be seen by a linear stability analysis, which yields (for real $\beta$)
\be
\dot y = -\lambda y, \quad\quad\quad
\lambda = \beta\cos x +\frac{4d}{1-\cos(2x)}. 
\ee
This is stable ($\lambda>0$), as long as $\beta/d\lesssim 5.196$. In fact, in the entire strip bounded by $\pm y_*$, the flow is directed towards the real manifold.

Let us now see how a complex coupling  changes the flow, see Fig.\ \ref{fig:class2} (right) for $\beta=1+0.5i$.
Even though the stable fixed points move away from $y=0$ [to $\pm (1.12, -0.158)$] and the unstable ones from $x=\pm\pi$ [to $\pm(2.79, -1.41)$], we note that in the region around $y\sim 0$ the flow is still directed to the real axis. This means that if initial conditions are chosen close to $y=0$, the complex stochastic process will take place in a strip around $y=0$. The probability distribution $P(x,y)$ will be strictly zero outside this strip and the theoretical foundation of the complex Langevin method is justified \cite{arXiv:0912.3360,arXiv:1101.3270}.
For larger (complex) $\beta$ the stable fixed points will move out further into the complex plane and the dynamics will eventually no longer be confined to a strip. The stability analysis as carried out here is then no longer applicable.

In the SU(3) case similar conclusions hold. The presence of the reduced  Haar measure is essential in stabilizing the real manifold (for nearly real dynamics, i.e.\ when $\beta_\rmI, \mu \ll 1$).
 Instead of showing four-dimensional flow diagrams, we look at the linear stability of the real manifold in the case that $h=0$ and $\beta$ is real. We write $\phi_a\to \phi^\rmR_a+i y_a$ and expand 
to linear order in $y_a$ ($a=1,2$). 
We find
\be
\left(
\begin{array}{c} \dot y_1 \\ \dot y_2 \end{array}
\right)
 = -A  \left(
\begin{array}{c} y_1 \\ y_2 \end{array}
\right),
\quad\quad\quad 
A=   
 \left(
  \begin{array}{cc} A_1 & A_o \\ A_o & A_2 \end{array}
\right).
\ee
The contributions from the action and the reduced Haar measure are denoted with $U$ resp.\ $H$ and we write $A=A^U+A^H$. We find
\be
A_a^U = 2   \beta \left[\cos(\phi_a^\rmR)+\cos(\phi_1^\rmR+\phi_2^\rmR)\right], 
\quad\quad\quad
A_o^U = 2   \beta \cos(\phi_1^\rmR+\phi_2^\rmR), 
\ee
 and 
\bea
\nn
A_a^H &=&  \half\csc^2\left(\frac{\phi_1^\rmR-\phi_2^\rmR}{2}\right)
+ 2 \csc^2\left(\frac{2\phi_a^\rmR+\phi_b^\rmR}{2}\right)
+ \half \csc^2\left(\frac{\phi_a^\rmR+2\phi_b^\rmR}{2}\right), \\
A_o^H &=&  -\half\csc^2\left(\frac{\phi_1^\rmR-\phi_2^\rmR}{2}\right)
+  \csc^2\left(\frac{2\phi_1^\rmR+\phi_2^\rmR}{2}\right)
+ \csc^2\left(\frac{\phi_1^\rmR+2\phi_2^\rmR}{2}\right),
\eea
where $b\neq a$ and $\csc z= 1/\sin z$.

It is easy to see that the eigenvalues of $A^H$ by itself are positive for all values of $\phi^\rmR_{1,2}$: the real manifold is stable when only this contribution to the drift is considered. On the other hand, the sign of the eigenvalues of $A^U$ depends on  $\phi^\rmR_{1,2}$, leading to unstable regions. In the combined linearized drift, with $A=A^U+A^H$, inspection shows that the real manifold is stable for $\beta \lesssim 1.75$, and (linearly) unstable for larger $\beta$.
We note that this bound covers most of the parameter space indicated in Fig.\ \ref{fig:hist}, except deep in the ordered phase, where $\beta$ can be larger (note that $\beta$ here refers to $\beta_{\rm eff}=6\beta|u|$ in Fig.\ \ref{fig:hist}). Here we remark that the linear stability analysis is probably too restrictive, since no problems were encountered in this region. In fact, in the ordered phase complex Langevin dynamics was seen to perform very well \cite{Aarts:2011zn}.

\subsection{Numerical results}

We have solved the complex Langevin equations numerically for a number of parameter values, using the two- and 
three-angle and the matrix formulations. Here we give a brief summary, illustrating the analytical findings discussed above.

 \begin{figure}[t]
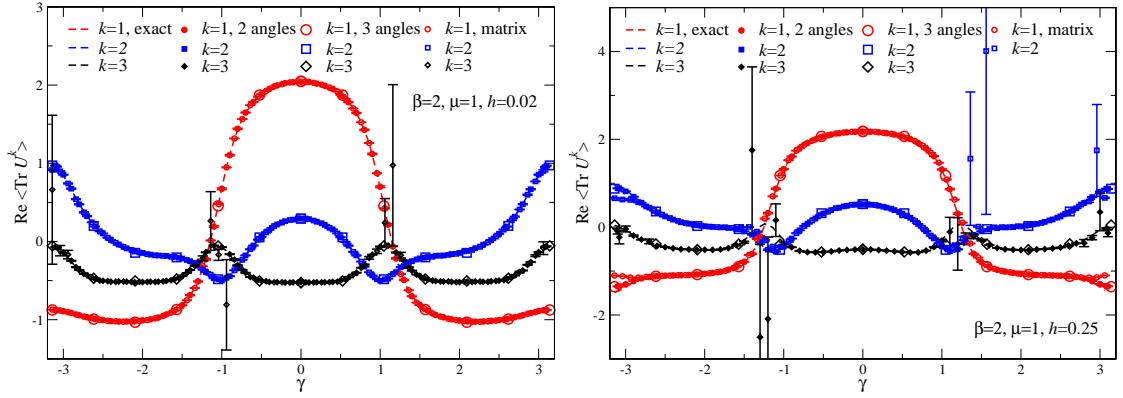

  \begin{center}
  \includegraphics[width=0.48\textwidth]{plot-b2-m1-h002-238angles-trUk-v2} 
   \includegraphics[width=0.48\textwidth]{plot-b2-m1-h025-238angles-trUk-v2} 
   \end{center}
 \caption{SU(3) one-link model: $\bra\Tr U^k\ket$ vs.\ $\gamma$ for $k=1,2,3$, at $\beta_{\rm eff}=2, \mu=1$ and $h=0.02$ (left) and 0.25 (right), using the two- and three-angle and the matrix formulations.  At $h=0.02$, agreement with the exact results is seen for all $\gamma$ in the two- and three-angle case; in the matrix case there is poor convergence at specific values of $\gamma$. At larger $h=0.25$, there is poor or wrong convergence also in the two- and three-angle case.
 }
 \label{fig:comp}
\end{figure}

Firstly, we find that, although they imply different processes,
 the two- and three-angle formulations give identical results (within the statistical error) and agree with the exact results for small values of $\beta$ and $h$, see e.g.\ Fig.\ \ref{fig:comp} (left) for $\beta=2, h=0.02, \mu=1$. Since $\mu\neq 0$, the dynamics does not take place on the real manifold but in the complexified configuration space. Nevertheless, the drift from the reduced Haar measure has a stabilizing effect, which constrains the dynamics. For larger values of the effective couplings $\beta_{1,2}(\mu)$ (which can  be achieved by increasing $\beta$, $h$ or $\mu$), we find that the dynamics may break down for a range of $\gamma$ values. This is illustrated in Fig.\ \ref{fig:comp} (right) for $\beta=2, h=0.25, \mu=1$. A breakdown can be characterized by large fluctuations (e.g.\ around $\gamma\sim \pm 1.25$) or by convergence to the wrong result (around $\gamma\sim\pm \pi$). 
This happens at parameter values where the dynamics is not clearly
dominated by a stable fixed point in the complexified field space.
Interestingly, when the one-link model is viewed as an effective model for the three-dimensional SU(3) spin model (or full QCD),  the combination of parameter values where the dynamics breaks down appears to be in  a region of parameter space which is  less relevant  for those; recall the discussion around  Fig.\ \ref{fig:hist}. Therefore this breakdown does not undermine the results obtained in the three-dimensional case.

In the matrix formulation, we observe that the dynamics is more subtle and can break down earlier. This process is distinct from the previous ones in that a different complexified field space is explored, with apparently more possibilities for unstable trajectories. This is visible in  Fig.\ \ref{fig:comp} already at $h=0.02$ for $k=3$ at specific $\gamma$ values and more so at larger $h=0.25$ (here the result at $k=3$ is not shown).
It is of course mandatory to control the dynamics in this formulation in the context of gauge theories and this can be partly achieved using gauge cooling \cite{Seiler:2012wz}.

  \begin{figure}
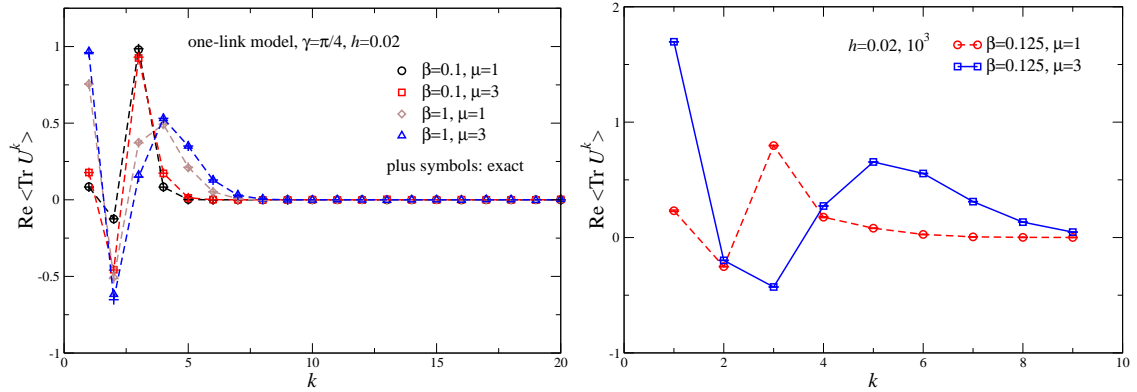

 \centerline{ 
  \includegraphics[width=0.488\textwidth]{plot_truk_g0_785_h0_02-1link-v2} 
  \includegraphics[width=0.48\textwidth]{plot_truk_b0_125_h0_02_10x10x10-ccc-v1} 
 }
 \caption{Expectation values $\bra \Tr U^k\ket$ as a function of $k$ for various values of $\beta$ and $\mu$ in the SU(3) one-link model (left) and in the three-dimensional model on a $10^3$ lattice. On the left, exact results are indicated with plus symbols; these can hardly be distinguished from the complex Langevin results.
 }
 \label{fig:truk}
\end{figure}

The mathematical foundation for complex Langevin dynamics to converge to the correct result is that the distribution in the complexified field space falls off rapidly. 
In ref.\ \cite{Aarts:2011zn} it was found that in the three-dimensional SU(3) spin model the distribution in the imaginary direction $\phi^\rmI$ drops exponentially, $P(\phi^\rmI) \sim e^{-b|\phi^\rmI|}$, with $b\sim 35-45$. 
This left open the question what happens to expectation values of the form $\bra\Tr U^k\ket$ with $k$ large. These observables contain terms of the form $e^{\pm k\phi^\rmI} \cos(k\phi^\rmR)$ and the presence of the rapidly oscillating cosines should be taken into consideration. This is demonstrated in Fig.\ \ref{fig:truk} where $\bra \Tr U^k\ket$ is shown as a function of $k$ in the one-link model (left) and in the three-dimensional model on a $10^3$ lattice (right). In the one-link model, exact results are shown as well;  agreement between the Langevin and exact results is observed.  We conclude therefore that the higher moments decrease rapidly and that the falloff of the distribution in the imaginary direction is sufficient to achieve this.

 Our conclusion from the numerical experiments with two and three angles is that the Langevin dynamics is controlled and yields the correct results  when the contribution from the reduced Haar measure is not overpowered by the contribution from the action itself. This is achieved by taking the effective couplings not too large. 
For larger $\beta$  or $h$ values however, the real manifold is no longer stable and this  implies problems for the complex Langevin process. 
Fortunately, this happens for parameter values in the effective one-link model, which do not seem to be relevant for the three-dimensional (or four-dimensional) original model.

Above the one-link model was used as an effective model for the
SU(3) spin model. However,  we can also see it as an effective model for 
lattice QCD in the heavy dense approximation, cf.\ refs.\ \cite{Aarts:2008rr,Seiler:2012wz,Fromm:2011qi,Fromm:2012eb}. In this case the 
effective coupling $\beta$ represents the contributions from the staples attached 
to a link, and instead of Eq.~(\ref{eq:sf}) the heavy dense 
determinant itself appears in the three-dimensional effective action:
\be
S_F = -  \sum_x \ln  \left[ \det \left(1 + h e^{\mu/T} {\cal P}_x\right)^2  \det \left(1 + h e^{-\mu/T} {\cal P}_x^{-1}\right)^2\right],
\ee
with $h =(2 \kappa)^{N_t}$ and ${\cal P}_\vecx^{(-1)}$ the (conjugate) Polyakov loops.
Using the determinant in the corresponding one-link model appears to have a stabilizing effect and the two- and three-angles simulations are even better behaved.


 \section{Stabilization via generalizations of the CLE}
 \label{sec:trafo}

A general possible route for stabilization uses the fact that there is a 
wide range of possibilities to modify the complex Langevin equation (CLE). Such a modification, 
if chosen appropriately, can improve the falloff of the equilibrium 
distribution, which was identified in refs.\ \cite{arXiv:0912.3360,arXiv:1101.3270} as the essential 
prerequisite for a correct CLE process. Some of those modifications, like 
kernels, were known already in the 1980s \cite{soderberg,okamoto,okano}, 
others such as variable transformations, were encountered just recently. 
In fact the reduction to the Cartan subgroup discussed in the previous 
section falls into this general category. In the present section we 
analyze the possible modifications more generally and find an equivalence 
between the approaches using a kernel and coordinate transformations.

The crucial point is that we apply the CLE only for averaging {\it 
holomorphic} functions. Different distributions in the complexified 
configuration space can give the same expectation values for those 
holomorphic functions; this is true for the probability density evolving 
under the Fokker-Planck equation as well as the stationary equilibrum 
distribution. In Appendix A we give general classes of processes and 
equilibrium measures which are equivalent in this sense. In addition there 
is the well-known freedom of using different processes having the same 
equilibrium measure.

Let us first describe the use of coordinate transformations. We have to 
distinguish two kinds of transformations:

\begin{itemize}
\item[(a)] transforming the CLE process, i.e.\ the description of the 
trajectories;

\item[(b)] transforming the variables in the functional integral.

\end{itemize}

The first possibility does not change anything essential, since it is only 
a transformation of the description; the second choice, however, leads to 
a modified process due to the appearance of a Jacobian. It turns out that 
after changing the description of this modified process back to the 
original coordinates one has in fact introduced a kernel.

In one variable this works as follows: let 
\be
x=x(u)
\ee 
be an invertible smooth variable transformation. To avoid confusion we 
rename the action $S(x)$ when considered as function of $u$ by
\be
\tilde S(u)\equiv S(x(u))\,.  
\ee
We have
\be
Z=\int dx \, e^{-S(x)}=\int du\, e^{-\tilde S_{\rm eff}(u)},
\ee
with 
\be
\tilde S_{\rm eff}(u)= \tilde S (u)-\ln J(u), \quad\quad\quad  J(u) = \frac{dx(u)}{du}.
\ee
The new drift becomes
\[ 
 K(u) = -\tilde S_{\rm eff}'(u) = -\tilde S'(u) +J'(u)/J(u),
\] 
and the new CLE
\be
\dot u= -\tilde S'(u) +J'(u)/J(u)+\eta(t). 
\ee
Let us first study the stability of the real manifold under small imaginary fluctuations in the original formulation. We write $x\to z=x+iy$ and linearize in $y$ to find
\be
\dot y  = -\lambda(x) y, \quad\quad\quad \lambda(x) = S^{\prime\prime}(x).
\ee
Only if $\lambda(x)>0$ for all $x$ is the real axis stable against complex perturbations.

By choosing a clever transformation one can stabilize a 
situation which is unstable due to a nonpositive $\lambda(x)$ or a repulsive fixed point.  
After the change of variables to $u$ and writing $u\to u+iv$, one finds,  to first order in $v$,
\be
\dot v  = -\tilde\lambda(u) v,
\ee 
where now
\be
 \tilde\lambda(u) = \tilde S^{\prime\prime}_{\rm eff}(u) = 
\tilde S^{\prime\prime}(u) - J^{\prime\prime}(u)/J(u) + \left[J^\prime(u)/J(u)\right]^2.
\ee
The contribution from the Jacobian may now stabilize the real manifold.

\begin{figure}
  \centerline{
    \includegraphics[width= 0.45\textwidth]{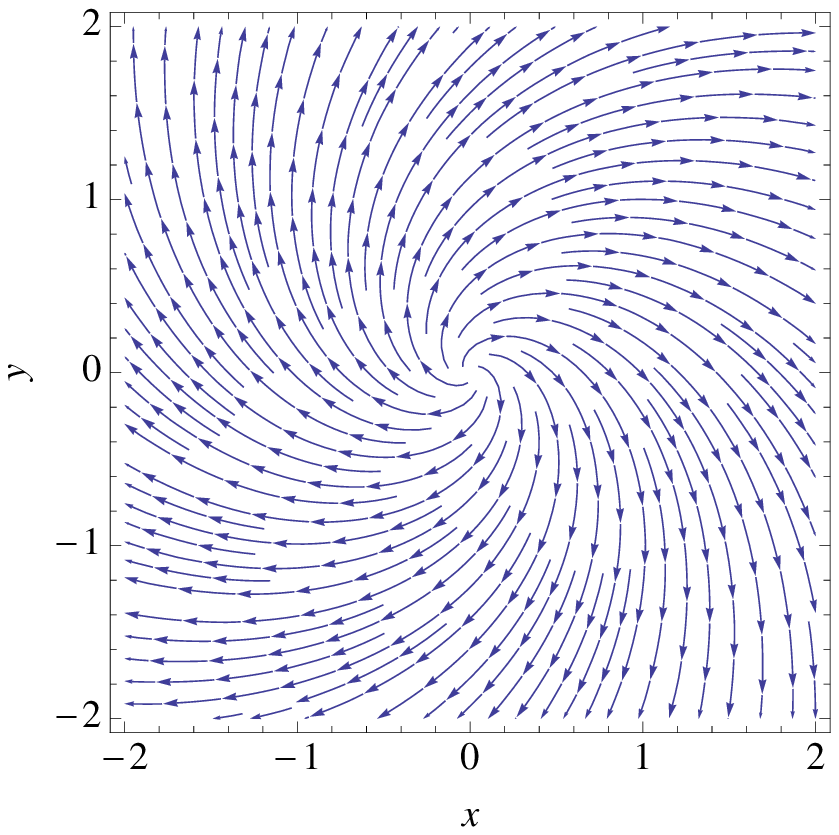}
     \includegraphics[width= 0.45\textwidth]{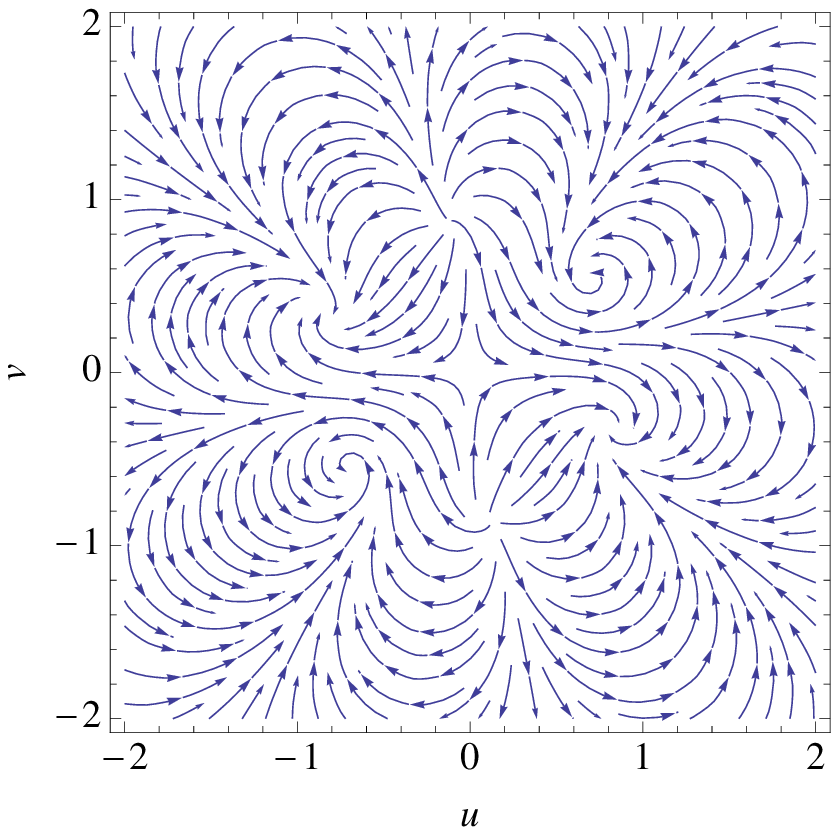}
  }
  \caption{Classical flow diagrams in the Gaussian model with 
$\sigma=-1+i$ in the original formulation (left) and after the coordinate 
transformation $x=u^3$ (right).
  }
\label{flowquad}
\end{figure}

 A striking example is given by the Gaussian model, with the action
\be
S=-\frac{1}{2}\sigma x^2, \quad\quad\quad \sigma = a+ib,
\ee
which is well defined as long as ${\rm Re}\, \sigma= a >0$. The origin 
is an attractive fixed point in this case. But for $a<0$ not only does the integral become ill-defined, 
after complexification also the fixed point at the origin becomes repulsive, see Fig.~\ref{flowquad} (left).

\begin{figure}
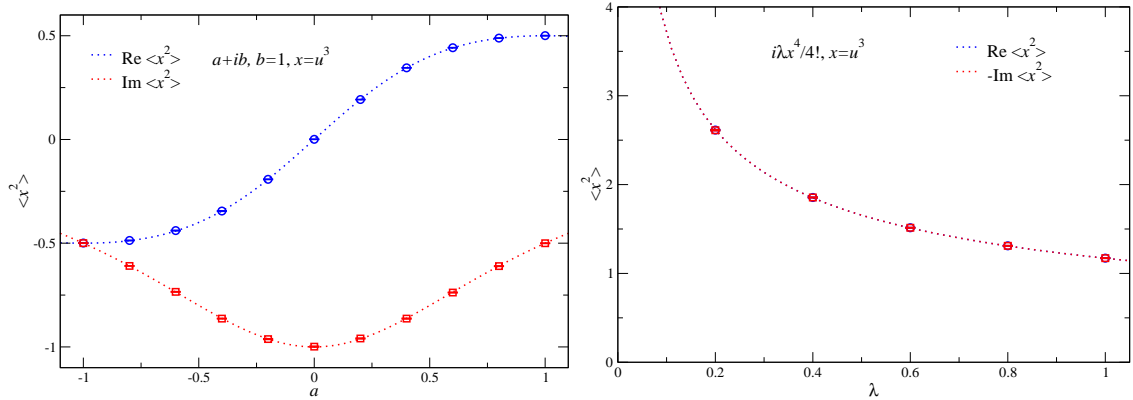

  \centerline{
     \includegraphics[width= 0.49\textwidth]{gaussian_a_b1}
     \includegraphics[width= 0.48\textwidth]{quartic_a0_b}
  }
\caption{Complex Langevin results for the  correlator $\bra x^2\ket$ in the Gaussian model with $\sigma=a+i$ as a function of $a$ (left) and in the quartic model as a function of $\lambda$ (right), both after the transformation $x=u^3$. The lines indicate the expected, analytically continued results.}
\label{fig:u3}
\end{figure}

A simple transformation $x(u)=u^n$ ($n$ odd) may remedy this situation, as shown in 
Fig.\ \ref{flowquad} (right) for $\sigma=-1+i$ and $n=3$:
while the additional term in the drift, $J'(u)/J(u) = (n-1)/u$, yields a singularity at the origin, it also partly stabilizes the real axis, since
\be
 \tilde\lambda(u) = \sigma n(2n-1) u^{2(n-1)} + \frac{n-1}{u^2},
\ee
and the second term is always positive. Most importantly, however, the change of variables leads to 
the appearance of stable fixed points in the 
complex plane. We find that a numerical solution of the CLE in $u$ converges  and yields for $\bra x^2\ket= 
\bra u^6 
\ket$
\be
\bra x^2\ket=\frac{1}{\sigma}= \frac{a+ib}{a^2+b^2},
\ee
even for $a\le 0$, see Fig.\ \ref{fig:u3} (left). This answer is correct in the sense that it is the 
analytic continuation of the result valid for $a>0$.

The same transformation also allows to evaluate the quartic `Minkowski' 
integral \cite{Berges:2005yt,Duncan:2012tc}
\be
Z = \int_{-\infty}^\infty dx\, \exp\left(-\frac{i\lambda}{4!} x^4\right),
\ee
defined by analytic continuation, giving
\be
\bra x^2\ket = \frac{2\sqrt{3}}
{\sqrt{\lambda}}\frac{\Gamma(\frac{3}{4})}{\Gamma(\frac{1}{4})}(1-i),
\ee
see Fig.\ \ref{fig:u3} (right).

\begin{figure}
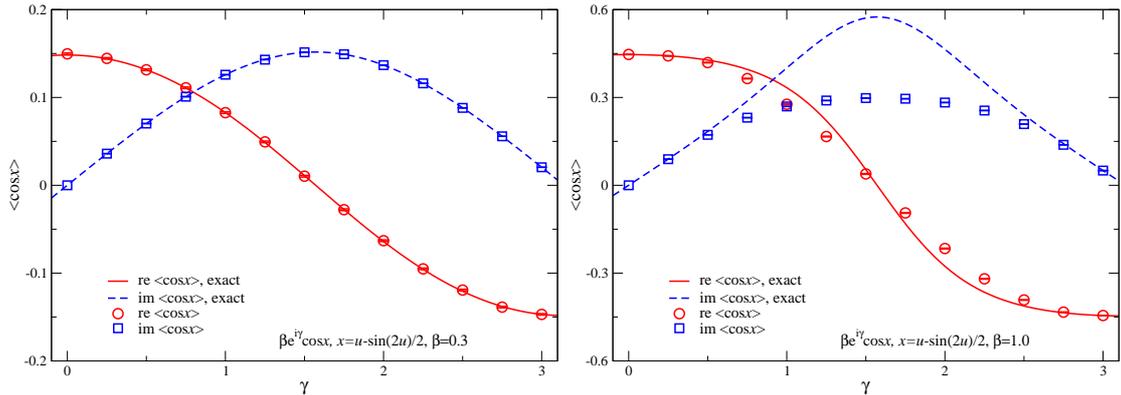

  \centerline{
     \includegraphics[width= 0.48\textwidth]{plot_cosx_u_b03_gamma}
     \includegraphics[width= 0.48\textwidth]{plot_cosx_u_b1_gamma}
  }
\caption{Complex Langevin results for the  correlator $\bra \cos x\ket$ in the U(1) model, with complex coupling $\beta e^{i\gamma}$  as a function of $\gamma$ at $\beta=0.3$ (left) and $\beta=1.0$ (right), after the transformation $x=u-\half\sin(2u)$.}
\label{fig:cosx}
\end{figure}

As the third example, we take a more realistic case, encountered in the previous section, namely 
\be
Z= \int_{-\pi}^\pi dx\, e^{\beta\cos x}   \quad\quad\quad (\beta\in \mathbb{C}).
\ee
We would like to stabilize this by a variable transformation that casts the integral in the form of an SU(2)-like integral, see Eq.\ (\ref{eq:usu}). This can be achieved by the change of variables, 
\be
x(u) = u-\half \sin(2u), \quad\quad\quad -\pi<u<\pi, 
\ee
since the Jacobian is
\be
J(u) = \frac{dx(u)}{du} = 2\sin^2 u.
\ee
Simulations in the new formulation yield results shown in Fig.\ \ref{fig:cosx}. The results are in line with the findings from above. For small $\beta=0.3$ (left), the Jacobian indeed has a stabilizing effect and the exact results are reproduced. For larger $\beta=1$ (right), the unstable contribution from the action takes over and exact results are only reproduced for $\gamma\sim 0,\pi$.  In the three-dimensional XY model, a variable change along these lines gave only partial success, unfortunately.

We now want to show that these transformations are equivalent to the 
introduction of a kernel. The CLE with a kernel can be written as
\be
\dot z= H^2(z) K(z)+ 2 H(z) H'(z) + H(z)\eta(t),
\label{kernelc}
\ee
with $z=x+iy$ and a holomorphic function $H$ (see refs.\ \cite{soderberg,okamoto, 
okano}, where $H^2$ is called the kernel). Separating real and imaginary 
parts, Eq.\ (\ref{kernelc}) becomes
\be
\dot x=  \hat K_x + {\rm Re}\,H \eta(t), 
\quad\quad\quad
\dot y=  \hat K_y  +{\rm Im}\, H \eta(t),
\ee
with 
\be
\hat K\equiv -H^2  \nabla_z S+\nabla_z H^2,
\quad\quad\quad
\hat K_x\equiv{\rm Re}\,\hat K\,
\quad\quad\quad
\hat K_y\equiv{\rm Im}\,\hat K.
\ee
Let us now change the description of the transformed process back to the 
original variables. This requires the use of  Ito calculus and the 
chain rule; one obtains the process 
\be
\dot z=\left[-\frac {1} {(u')^{2}} S'(z) - \frac {2 u''} {(u')^3}\right] 
+ (u')^{-1} \eta(t),
\ee
This can be recognized as the original process modified by a kernel using
$H=1/u' = J$. 
The fact that $H$ strictly speaking is not a holomorphic kernel because 
of its singularity at the origin is of no consequence in the cases 
discussed before, because the equilibrium distribution vanishes at 
this point. 

The reduction of the SU(2) and SU(3) effective models to the Cartan 
subgroup discussed in Section 3 is an example of singular variable 
transformation; the same is true about the relation between the $N$ angle 
and the $(N-1)$ angle formalisms discussed there. In Appendix A these 
issues are discussed in a more systematic way.

\section{Conclusion}
\label{sec:conclusion}

Complex Langevin dynamics can handle theories with a complex action, even when the sign problem is severe, but success is not guaranteed. In this paper we contrasted the apparent success in the three-dimensional SU(3) spin model, an effective model for QCD at nonzero chemical potential, with the observed failure in the three-dimensional XY model in the disordered phase, by constructing and analyzing effective one-link models.

We found that a crucial role is played by the presence of the nontrivial Haar measure in the nonabelian case. The contribution from the reduced Haar measure to the drift constrains the dynamics in the complexified field space: it is purely restoring and directed towards attractive fixed points on the real manifold. The presence of a singularity in the drift at the origin does not cause problems. On the other hand, the drift  from the action yields repulsive fixed points, which can be located on or close to the real manifold and are responsible for large excursions in the complexified field space. As we have shown earlier, an uncontrolled exploration of the enlarged field space results in insufficient falloff of the equilibrium distribution,  which causes the formal justification of the complex Langevin equation to break down  and also leads to 
wrong results in practice.

In the nonabelian case, the contributions from the reduced Haar measure and action balance each other, resulting in a controlled Langevin process. We showed that the real manifold is stable against small complex fluctuations.  This ensures
analyticity of the observables in e.g.\ $\mu^2$.
By  increasing the importance of the contribution from the action, the Langevin process can be made to break down. However, the parameter values in the effective one-link models  for which this happens do not seem to be relevant for the three-dimensional SU(3) spin model.
The results of this paper provide further justification for the conclusion drawn in ref.\ \cite{Aarts:2011zn}, in which complex Langevin was argued to be applicable  by carrying out a series of assessments, without relying on knowledge of the  `exact' result.

In the (abelian) XY model the Haar measure is trivial: hence there is no restoring component  and the real manifold is unstable against small complex fluctuations. Since these fluctuations are introduced as soon as $\mu\neq 0$, simulations when $\mu\to 0$ and $\mu = 0$ do not agree. This is indeed what has been observed in the disordered phase of three-dimensional XY model and is now fully explained.
Deep in the ordered phase, the theory becomes effectively real and there is no problem for complex Langevin dynamics.

The important lesson is that it is desirable for the real manifold to be stable against complex fluctuations in order for complex Langevin dynamics to work (even though strictly speaking  it is neither sufficient nor necessary).
 This provides a useful benchmark to address,  given that there is a wide range of options to modify the complex Langevin approach (some of them were discussed here).
This is especially relevant for gauge theories, where the complexification leads to an extension from SU($N$) to SL($N, \mathbb{C}$). Here it is possible to stabilize the real group manifold via gauge cooling, at least partly \cite{Seiler:2012wz}.


 \vspace*{0.5cm}
 \noindent
 {\bf Acknowledgments.} \\
We thank MPI M\"unich (GA, FAJ, DS and IOS) and Swansea University (JMP, ES, DS and IOS) for its hospitality. GA, FAJ, ES, DS and IOS also thank the Department of Energy's Institute for Nuclear Theory at the University of Washington for its hospitality and the Department of Energy for partial support during stages of this work. The work of GA and FAJ is supported by STFC and carried out as part of the UKQCD collaboration and the DiRAC Facility jointly funded by STFC, the Large Facilities Capital Fund of BIS and Swansea University. FAJ is supported by the
European Union under Grant Agreement number 238353 (ITN STRONGnet).  


\appendix
\section{Generalizations of Langevin dynamics}
\label{sec:app}

In this appendix we investigate systematically possible generalizations of 
complex Langevin (CL) dynamics that are at least formally equivalent 
to the original one. For the formal equivalence (i.e. ignoring problems of 
slow falloff and possible boundary terms, see refs.\ \cite{arXiv:0912.3360,arXiv:1101.3270}) it is 
sufficient to verify that the (complex) Fokker-Planck (FP) operator has 
$\exp(-S)$ as a zero mode. In refs.\ \cite{arXiv:0912.3360,arXiv:1101.3270}  it is also explained how the 
equilibrium distribution of the real FP operator acting on functions of 
the complexified configuration space becomes, when evaluated on 
holomorphic observables, equivalent to that zero mode.

\subsection{Nonuniqueness of measures for holomorphic observables}

For simplicity we consider the `flat' case of $\R^N$. The goal of the 
complex Langevin equation (CLE) is to find a probability measure on $\C^N$,
\be
P(x,y) dx dy  \quad\quad\quad x,y\in\R^N,
\ee
which for holomorphic observables $\cO$ is equivalent to a complex measure 
on $\R^N$,
\be
\rho(x) dx,
\ee
in the sense that for all $\cO$,
\be
\int_{\C^N} dx dy \, P(x,y) \cO(x+iy) = \int_{\R^N} dx\,  \rho(x)\cO(x).
\label{equiv}
\ee
The question of the existence of such a measure $P$ was already discussed 
by  Weingarten \cite{weingarten}, but here we are interested in the 
nonuniqueness.

Clearly $P$ is not uniquely determined by the requirement (\ref{equiv}). 
First one has to recognize that a precise answer to this nonuniqueness 
depends on specifying the space of holomorphic observables for which 
Eq.\ (\ref{equiv}) is supposed to hold. We consider spaces $\cH_f$ defined in 
terms of some growth condition
\be
\sup \sigma(x,y)|\cO(x,y)|<C<\infty ,
\ee
for some positive weight function $\sigma$. Examples for $f$ are 
$f=\exp((x^2+y^2)^\alpha)$, with $x^2=x_1^2+\ldots x_N^2$, etc.

Obviously $P_1$ and $P_2$ are equivalent in the sense above if and only if 
for all $\cO\in \cH_f$ $Q\equiv P_1-P_2$ satifies 
\be 
\int_{\C^N} dx dy \, Q(x,y) \cO(x+iy)=0. 
\label{requ} 
\ee 
In other words, $Q$ has to be orthogonal to the intersections of the 
kernels of the Cauchy-Riemann operators
\be
C_j\equiv \partial_{x_j}+i \partial_{y_j}.
\ee
Ignoring possible convergence problems or boundary terms, Eq.\ (\ref{requ}) 
seems to say only that $Qdx dy$ is a signed measure satisfying
\be
Q(x,y)=\sum_{j=1}^N C_j H_j(x,y),
\label{modmeas}
\ee
where the $H_j$ are some complex-valued functions (of bounded variation), 
because rolling the CR operators $C_j$ over to act on the observables 
would yield zero. Those conditions are satified if there are real functions 
$G_j(x,y)$ satisfying 
\be
Q(x,y)=\sum_{j=1}^N \Delta_j G_j(x,y)
\ee
with $\Delta_j= \partial_{x_j}^2+\partial_{y_j}^2$. 

The convergence problems have to be taken seriously; we are only sure 
that they are absent if $Q\cO$ is absolutely integrable. Depending on the 
class of holomorphic observables chosen, one has to demand corresponding 
decay properties of the $G_j$.

To make use of Eq.\ (\ref{modmeas}) for a cure of convergence to the wrong 
limit, one would need to do the following: first find a $Q$ that is such 
that it improves the decay of $P+Q$ at infinity (this is of course only 
possible if $Q$ shows that same rate of decay as $P$), then find a 
modified CL process having $P+Q$ as its equilibrium measure.

\subsection{Neutral modification of the CL process}
 
Next we ask how one can modify the CL process without changing the 
evolution of holomorphic observables under the Langevin evolution; we call 
this a neutral modification. If the Langevin operator is given by 
\be L=[\nabla_x+K_x] \nabla_x + K_y \nabla_y, 
\ee 
we want to see if we can replace it by an operator of the general form 
\be L'\equiv L+\sum_j F_j(x,y)^2 \partial_{x_j}^2+\sum_j G_j(x,y) ^2 
\partial_{y_j}^2 + R_x \cdot \nabla_x+ R_y \cdot \nabla_y, 
\ee 
with coefficient functions $F_j,G_j, R_x,R_y$ ($F,G\ge 0$) in such a way 
that when applied to holomorphic observables $\cO$, $L'$ becomes equal 
to $L$, because $\partial_{y_j}\cO= i \partial_{x_j}\cO$ ($j=1,\ldots N$).

It is not hard to see that this requires
(1) $G_j^2=F_j^2$ ($j=1,\ldots N$);
(2)  $R_x=R_y=0$.
So the modifying operator is of the form
\be
L_m= \sum_j F^2_j(\partial_{x_j}^2+\partial_{y_j}^2),
\ee
corresponding to the modified CL process
\be
\dot x_j = K_{x_j} +\sqrt{1+ F_j^2}\,\eta_j(t),
\quad\quad\quad
\dot y_j = K_{y_j} + F_j\,\eta_j(t).
\ee   
A well-known special case arises by choosing 
\be
F_j(x,y)=N_I,\quad j=1,\ldots N,
\ee
which we refer to as `complex diffusion'. Here we have discovered a 
generalization: `configuration-dependent complex diffusion'. But it is not 
clear how much this can help in solving the problems of the CLE, in view 
of the fact that it was found that introducing `imaginary diffusion' (or 
complex noise) actually tends to create problems by leading to slow 
decay in the imaginary directions and convergence to the wrong limit 
\cite{arXiv:0912.3360,arXiv:1101.3270}. So $F_j$ should at least be chosen in such a way that it 
vanishes sufficiently fast for 
$|y_j|\to\infty$, $y=1,\ldots, N$. 

\subsection{Nonneutral modifications of the CLE: holomorphic (matrix) 
kernels}
\label{hol}

A way of modifying the CL process that was discussed already in the 1980s is 
the introduction of a so-called kernel \cite{soderberg, okamoto,okano}. It 
is nonneutral in the sense that it does not preserve the Langevin 
evolution of holomorphic observables (but is supposed to preserve the 
long-time averages of holomorphic observables). For one dimension it 
amounts to the following:

Let $H$ be holomorphic on $\C$ ($H$ corresponds to what is called $\sqrt{K}$ 
in the older literature). We denote
\be
{\rm Re}\, H\equiv R,\quad\quad\quad {\rm Im}\, H\equiv I.
\ee
The corresponding CLE becomes
\be
\dot x = \hat K_x+ R \eta(t), 
\quad\quad\quad
\dot y = \hat K_y+ I \eta(t),
\ee
where
\be
\hat K\equiv -H^2 S'+2HH', 
\quad\quad\quad
\hat K_x\equiv{\rm Re}\,\hat K,
\quad\quad\quad
\hat K_y\equiv{\rm Im}\,\hat K. 
\ee
It is easy to find choices for the kernel that eliminate the drift force 
altogether or change its sign, but the resulting process suffer from 
steeply rising diffusion coefficients and dubious stability.

A special case that does not suffer from these problems is a constant 
(complex) kernel $H$, with ${\rm Re}\, H^2>0$. In this case the kernel just 
amounts to multiplying the time coordinate by a factor $H^2$ and there are 
no consistency problems. This simple device has been shown to improve the 
situation in certain cases with quadratic action in 
refs.\cite{okamoto,okano}; the stabilization occurring there is actually 
equivalent to the one discussed in Section 4.

For more than one variable there is a more general way of introducing 
kernels which we call matrix kernels. In short, take a holomorphic
$H_{jk}(x_1+iy_1,\cdots, x_n+iy_n)$, then
\be
\dot x_j = \hat K_{x,j}+ {\rm Re}\, \sum_k H_{jk}\eta_k(t),
\quad\quad\quad
\dot y_j =\hat K_{y,j}\notag +{\rm Im}\,\sum_k H_{jk}\eta_k(t), 
\ee
where
\begin{align}
&\hat K_j\equiv \sum_k \left\{-(H^T H)_{kj} (\nabla_{z_k} S)+
 \nabla_{z_k} (H^T H)_{jk}\right\}, \notag\\
&\hat K_{x,j}\equiv{\rm Re}\,\hat K_j, 
\quad\quad\quad
\hat K_{y,j}\equiv{\rm Im}\,\hat K_j.
\end{align}

To see that this is (formally) correct, note that
the complex Fokker-Planck operator is
\be
L^T_H= \sum_{k,j}\left\{\nabla_j (H^T H)_{kj}\left[\nabla_k+(\nabla_k S)\right]\right\}.
\label{fpk}
\ee
It is manifest that $\exp(-S)$ is a (hopefully unique) zero mode of $L^T$.

We remark that we can actually combine the two previous modifications 
(kernel and imaginary diffusion). For simplicity we only describe the case 
of one dimension: let again $H$ be a holomorphic function and $F$ a real 
valued function on $\cM_c$. Then the modified Langevin operator
\be
L_H = \left( (R^2+F^2)\nabla_x+\hat K_x\right)\nabla_x
+\left((I^2+F^2)\nabla_y+\hat K_y\right) \nabla_y +2RI\nabla_x\nabla_y
\ee
reduces on holomorphic observables to 
\be
\tilde L_{H,F}= H^2\Delta_z +\hat K \nabla_z,
\ee
as before.

We have to find a process corresponding to these operators. This problem 
does not have a unique solution, but the following stochastic differential 
equation turns out to give the right $L_{H,F}$:
\bea
\dot x&=& \hat K_x+ 
\frac{R\sqrt{R^2+I^2+F^2}}{\sqrt{R^2+I^2}}\eta_1(t)-\frac{FI}{\sqrt{R^2+I^2}}
\eta_2(t),
\notag\\
\dot y&=& \hat K_y+\frac 
{I\sqrt{R^2+I^2+F^2}}{\sqrt{R^2+I^2}}\eta_1(t)
+\frac{FR}{\sqrt{R^2+I^2}}\eta_2(t)\,,
\eea
where $\eta_1, \eta_2$ are two independent white noises. Note that these equations reduce to the previous ones for the 
special case $F=0$; for $F=N_I$ and $R=1,\,I=0$ one recovers the standard 
complex noise.

\subsection{Nonneutral modifications of the CLE: Coordinate 
transformations}

As remarked in Section 4, coordinate transformations are closely related 
to kernels. In fact they lead to a subset of the general matrix kernels 
discussed above, but since they are sometimes easier to handle we discuss 
them in some detail. To avoid heavy notation, we treat real actions and 
comment about the complex case in Subsection A.4.6. 

\subsubsection {General remarks on coordinate transformations}

As stated in Section 4, we have to distinguish two types of coordinate
transformations:

(I) Transforming the descriptions of the trajectories;

(II) Transforming the functional integral.

Choice (I) does not change the process, only its description, but choice 
(II) is nontrivial. The difference can be seen clearly in a simple linear 
one-dimensional example: let
\be
u=ax\,\,(a\neq 0)
\ee 
and consider the real Langevin process corresponding to the action 
$S(x)=S(u/a)\equiv T(u)$:
\be
\dot x=-S'(x) + \eta(t). 
\ee  
Then (I) means 
\be
({\rm I}):\quad \dot u=-a^2 T'(u) +a \eta(t),
\ee
whereas (II) means replacing the measure proportional to $\exp(-S(x))dx$ 
by the transformed one proportional to $\exp(-T(u))du$ and setting up a 
Langevin process that has the transformed measure as its equilibrium 
measure. We find
\be
({\rm II}):\quad \dot u=-T'(u)+\eta(t).
\ee
So in this simple example the difference between (I) and (II) is just a 
different time scale.

\subsubsection{Nonsingular (invertible) linear transformations}

We now consider a less trivial case of a Langevin process in $\R^N$, 
using vector notation:  
\be
\dot x= -\nabla S(x) + \eta(t), 
\ee
and the transformation by a nonsingular matrix $A$ 
\be
u=A\,x, \quad\quad\quad T(u)\equiv S(A^{-1}\,u).
\ee
Scheme (I) yields, simply by acting with $A$ on both sides of the 
LE,
\be
\dot u=-A\nabla_x S(x)+ A d\eta \,\quad {\rm for} 
\quad u=A\,x.
\ee
To re-express the drift in terms of $T(u)$, we use the chain rule
\be
\nabla_x T(A x)=A^T \nabla_u T(u)\,\quad {\rm for} 
\quad u=A\,x, 
\ee 
which leads to
\be
({\rm I}):\quad \dot u=-AA^T \nabla T(u) +A \eta(t),
\ee
whereas (II) gives
\be
({\rm II}): \quad \dot u = -\nabla T(u) +\eta(t).
\ee
The two processes now no longer differ just by different times scales. 
It is also instructive to compare the Fokker-Planck operators for the two 
processes. In matrix notation
\be 
({\rm I}):\quad L^T= \nabla A A^T (\nabla+ (\nabla T)) 
\ee
and 
\be
({\rm II}):\quad L^T=\nabla(\nabla+ (\nabla T)).
\ee
It obvious that both operators have the same equilibrium solution. 
The transition from scheme (II) to scheme (I) can be viewed as the 
introduction of a constant `matrix kernel' $A$; in this sense it is more 
general than the standard introduction of scalar kernels.    

\subsubsection{Singular (noninvertible) linear transformations}
\label{singlin}

If the action has a certain symmetry, noninvertible linear transformations 
can be used to reduce the number of variables to a subset; conversely one 
can use them to introduce a redundant set of variables, which is sometimes 
useful.

Here we show explicitly how this can be used to justify the process on the 
Cartan subgroup of $SU(N)$, using $N$ instead of $N-1$ angles, as 
discussed in sec.\ \ref{sec:sun}. We have to consider singular linear 
transformations $A$ that map $\R^N$ into a lower dimensional space $\R^M$ 
(with $M<N$): 
\be 
\label{eq:uAx}
u=A\,x. 
\ee 
$A$ has a null space $V_o\subset \R^N$ and 
an orthogonal complement $V_\perp$. We assume that the action is invariant 
under shifts in the $V_0$ directions, i.e. 
\be 
S(x+u)=S(x) \quad {\rm whenever} \quad u\in V_0. 
\ee 
Then the drift $\nabla S$ will leave $V_\perp$ invariant.

Specializing now to the case of SU(3) and going from the 3 angle to the 2 
angle formulation,  we may think of factorizing $A$ as
\be 
A=QB, 
\ee 
where $B$ is a singular symmetric $N\times N$ matrix annihilating $V_0$ 
and mapping $V_\perp$ onto itself. $Q$ is then a one to one mapping of 
$V_0$ onto $\R^M$.

In the case of Eq.\ (\ref{eqz}), with $x$ resp.\ $u$ in Eq.\ (\ref{eq:uAx}) denoted with $z$ resp.\ $\phi$ in  sec.\ \ref{sec:sun},  $B$ maps the $z$'s onto the subspace with 
$\phi_1+\phi_2+\phi_3=0$ in terms of the 3 angles $\phi_1,\phi_2,\phi_3$, 
so we have 
\be
B=\frac{1}{3}
\left(
\begin{matrix}
\ \ 2 &-1&  -1\\
-1&\ \ 2& -1\\
-1&-1 &\ \ 2
\end{matrix}
\right),
\ee 
and $Q$ maps onto the two independent angles $\phi_1, \phi_2$, i.e.
\be
Q=
\left(
\begin{matrix}
 1 & 0 & 0\\
 0 & 1 & 0
\end{matrix}
\right),
\ee
so finally
\be
A=QB=\frac{1}{3} 
\left(
\begin{matrix}
\ \  2 & -1 & -1\\
 -1 &\ \  2 & -1
\end{matrix}
\right).
\ee 

The LE for the variables $x\in\R^N$ is 
\be
\dot x=-\nabla S(x)  + \eta(t)\,,
\ee
where $\eta$ consists of $N$ independent white noises, hence does {\it 
not} leave the subspace $V_0$ invariant. The remainder of the analysis is 
almost identical to the one for nonsingular $A$: the LE for $u=A\,x$ 
(scheme (I)) can at first be written as
\be
\dot u=-A\nabla_x S(x) + A \eta(t)\,,
\ee
but we have to re-express the drift $-\nabla S(x)$ in terms of the 
transformed variables $u$. First we define
\be
T(u)\equiv S(x)\quad {\rm whenever} \quad  u=A\,x,
\ee
which is well defined because of the invariance of $S(x)$.  By the 
chain rule again
\be
\nabla_x T(A\,x)= A^T\nabla_u T(u),
\ee
so the LE for $u$ becomes
\be
({\rm I}):\quad \dot u = - AA^T \nabla_u T(u) + A \eta(t)\,.
\label{leu}
\ee

In scheme (II), on the other hand, the LE for $ u$ is simply
\be
({\rm II}):\quad \dot u=\nabla_u T(u) +\eta(t)\,,
\ee
where $\eta(t)$ consists now of $M$ independent white noises. To see to 
which extent the two schemes are equivalent, let us look at the 
corresponding FP operators:
\be
({\rm I}): \quad L^T_{\rm I} = \nabla_u^T AA^T  (\nabla_u+ (\nabla_u T))
\ee
and 
\be
({\rm II}): \quad L^T_{\rm II}= \nabla_u (\nabla_u+ (\nabla_u T))\,.
\ee
Again it is obvious that both schemes lead again to the same equilibrium 
measures, but clearly the processes are quite different.

\subsubsection{Invertible nonlinear coordinate transformations}

The treatment of nonlinear coordinate transformations is quite similar to 
the linear case, except that one has to use Ito's calculus (see for 
instance ref.\ \cite{friedman}) in some places. Again we have to consider 
both schemes (I) and (II). We first consider invertible transformations of 
$\R^N$ and comment in the next subsection about singular (noninvertible) 
ones.

Let $f: \R^N\to \R^N$ be an invertible and continuously 
differentiable coordinate transformation. If (in vector notation)
\be 
u=f(x)
\ee
and $x(t)$ evolves according to the Langevin equation
\be
\dot x =-\nabla S(x)+\eta(t),
\ee
we define again
\be
T(u)\equiv S(f^{-1}(u)).
\ee
Then by the Ito calculus, $u$ evolves according to the LE
\be
\dot u_i=- \sum_j \left\{\left[\frac {\partial u_i}{\partial x_j}
\frac{\partial S} {\partial x_j}+\Delta_x u_j\right] + 
\frac{\partial u_i}{\partial x_j} \eta_j(t)\right\}.
\ee
To express this LE in terms of $u$ and $T(u)$ we again use the 
chain rule
\be
\frac{\partial S(u)} {\partial x_j}=
\sum_r \frac{\partial u_r}{\partial x_j} 
\frac{\partial T(u(x))}{\partial u_r}.  
\ee
If we define the matrix $A$ by
\be
A_{ij}= \frac {\partial u_i} {\partial x_j},  
\label{matrix}
\ee
the LE for $u$ becomes in matrix notation
\be
({\rm I}):\quad \dot u=-AA^T  \nabla_u T(u) + \Delta_x u  + 
A \eta(t),
\ee
which clearly reduces to Eq.\ (\ref{leu}) in the linear case.  

In scheme (II) the transformation of the functional integral involves the 
Jacobian 
\be
J=\frac{\partial(x)}{\partial (u)}=\det A^{-1},
\ee
so the action $T(u)$ receives an extra term 
\be
T_J\equiv -\ln J = \ln \det A.
\ee
With 
\be
T_{\rm eff}\equiv T+T_J = T +\ln\det A,
\ee
we get the Langevin equation for scheme (II) 
\be
({\rm II}): \quad \dot u=-\nabla_u T_{\rm eff}(u) + \eta(t)\,,
\ee
with the $\eta(t)$ denoting again $N$ independent white noises. 

To see that the two schemes are equivalent as far as their equilibrium 
measures are concerned, we again compare the Langevin and FP operators of 
the two schemes. We first note a general fact (see for instance ref.\ 
\cite{friedman}): if
\be
\dot x =K(x)  + A(x) \eta(t), 
\ee
then by the Ito calculus 
\be
L= \sum_{i,k} (A A^T)_{ik}(x) \partial_i\partial_k+
K(x)\cdot \nabla
\ee
and 
\be
L^T=\sum_{i,k} \partial_i\partial_k (A A^T)_{ik}(x)- 
\nabla \cdot K(x). 
\ee
Applying this to our two Langevin equations above we find for scheme (I)
\be
({\rm I}):\quad L^T_I= \sum_{i,k}\frac{\partial}{\partial u_i}
\frac{\partial}{\partial u_k} (A A^T)_{ik} 
-\sum_i\frac{\partial}{\partial u_i}\Delta_x u_i
+\sum_{i,k} \frac{\partial}{\partial u_i} (A A^T)_{ik} \frac{\partial 
T}{\partial u_k} 
\ee
and for scheme (II)
\be
({\rm II}):\quad L^T_{\rm II}= \nabla_u^T \left[\nabla_u+
\left(\nabla_u T_{\rm eff}(u)\right)\right].
\ee
We claim that the FP operator in scheme (I) can be rewritten as
\be
({\rm I}):\quad L^T_{\rm I}= \nabla^T_u  A A^T\left[\nabla_u+(\left[\nabla_u 
T_{\rm eff}(u)\right)\right],
\ee
again making the equivalence of the two schemes obvious. To prove our 
claim we only have to show that
\be
\sum _k (AA^T)_{ik} \frac{\partial}{\partial u_k}  \ln \det A =
-\Delta_x u_i+ \sum_k\frac{\partial}{\partial u_k}(AA^T)_{ik};
\label{chain2}
\ee
this follows again using the chain rule:
\be
\sum_k\frac{\partial}{\partial u_k}(AA^T)_{ik}=
\sum_{k,l,r} \frac{\partial x_l}{\partial u_k} 
\frac{\partial}{\partial x_l}
\left( \frac{\partial u_i}{\partial x_r}
\frac{\partial u_k}{\partial x_r}\right).
\ee
The last expression is equal to
\be
\sum_{k,l,r}\left\{\frac{\partial x_l}{\partial u_k} 
\frac{\partial^2 u_i}{\partial x_l \partial x_r} 
\frac{\partial u_k}{\partial x_r} +
\frac{\partial x_l}{\partial u_k}  \frac{\partial u_i}{\partial x_r}
\frac{\partial^2 u_k}{\partial x_l \partial x_r} \right\}; 
\ee
here the first term simplifies to $\Delta_x u_i$, whereas the second term 
equals 
\be
\sum_k (AA^T)_{ik} \frac{\partial}{\partial u_i}  \ln \det A,
\ee
so Eq.\ (\ref{chain2}) is correct.

As in the linear case the transition from scheme (II) to scheme (I) can be 
viewed as the introduction of a `matrix kernel' $A$ (see Eq.\ \ref{matrix}), 
which is a generalization of the standard introduction of scalar kernels.
Again we can transform scheme (II) back to the original coordinates using 
scheme (I) and obtain a process with a matrix kernel $A^{-1}$.

But note that the matrix kernels obtained here are not as general as 
before, since the matrix $A^{-1}_{ij}=\partial x_i/\partial u_j $  has to 
satisfy an integrability condition, which was not required of the matrix 
$H$ in sec.\ \ref{hol}.

\subsubsection{Singular (noninvertible) nonlinear transformations}

As in the case of singular linear transformations, in this case we have to 
require some additional symmetries of the action. A simple example is the 
action 
\be
S=-\beta\, \tr U,
\ee
with $U$ an element of SU($N$). The symmetry in question is the invariance 
of $S$ under similarity transformations
\be
U \mapsto V^{-1}UV,
\ee
which allows the transition to the Cartan subgroup of SU($N$). The map 
from the $N^2-1$ parameters of SU($N$) to the $N-1$ Cartan angles is clearly 
a singular noninvertible transformation.

More generally, in a lattice gauge model gauge invariance allows to reduce 
the process to one remaining in a lower-dimensional gauge-fixed 
configuration space.

\subsubsection{Remark on complex actions}
 
If $S(x)$ is complex, at least on the formal level (ignoring possible 
problems with boundary terms under integration by parts) everything goes 
through as before.
  
The formal proof of correctness given in refs.\ \cite{arXiv:0912.3360,arXiv:1101.3270}  only uses the 
fact that the complex FP operator annihilates the complex measure 
$\exp(-S) dx$. This is manifestly the case for the most general case of a 
holomorphic matrix kernel, see Eq.\ (\ref{fpk}).

Here in principle we have to limit ourselves to transformations that are
bi-holomorphic maps from $z= x + i y$ to $ w = u+iv$. In practice it 
seems that this restriction is too severe and certain singularities can be 
tolerated -- provided we are lucky and the equilibrium density vanishes at 
those points. The formal proof of equivalence again only uses the 
equivalence of the complex Fokker-Planck operators -- the Langevin 
evolutions in the 0 space are again different in the two schemes.

\end{document}